\documentclass[sigconf,authorversion,nonacm]{acmart}

\usepackage{microtype}
\usepackage{graphicx}
\usepackage{subcaption}
\usepackage{booktabs} % for professional tables
\usepackage{natbib} % bibtex

\usepackage{hyperref}
\usepackage{amsmath}
\usepackage{multirow}
\usepackage{url}
\usepackage{amssymb}
\AtBeginDocument{%
  \providecommand\BibTeX{{%
    \normalfont B\kern-0.5em{\scshape i\kern-0.25em b}\kern-0.8em\TeX}}}

\begin{document}

\title{Bayesian Time Varying Coefficient Model with Applications to Marketing Mix Modeling}

\author{Edwin Ng}
\affiliation{%
  \institution{Uber Technologies, Inc}
  \country{} 
   \city{}
}
\email{edwinng@uber.com}
\author{Zhishi Wang}
\affiliation{%
  \institution{Uber Technologies, Inc}
    \country{} 
    \city{}
}
\email{zhishiw@uber.com}
\author{Athena Dai}
\affiliation{%
  \institution{Uber Technologies, Inc}
    \country{} 
    \city{}
}
\email{athena.dai@uber.com}


\begin{abstract}
Both Bayesian and varying coefficient models are very useful tools in practice as they can be used to model parameter heterogeneity in a generalizable way. Motivated by the need of enhancing Marketing Mix Modeling (MMM) at Uber, we propose a Bayesian Time Varying Coefficient (BTVC) model, equipped with a hierarchical Bayesian structure. This model differs from other time-varying coefficient models by weighting coefficients over a set of local latent variables that follow specific probabilistic distributions. Stochastic Variational Inference (SVI) is used to approximate the posteriors of latent variables and dynamic coefficients. The proposed model also helps address many challenges faced by traditional MMM approaches. We used simulations as well as real-world marketing datasets to demonstrate our model’s superior performance in terms of both accuracy and interpretability.
\end{abstract}


%\begin{CCSXML}
%<ccs2012>
%   <concept>
%       <concept_id>10003752.10003753.10003757</concept_id>
%       <concept_desc>Theory of computation~Probabilistic computation</concept_desc>
%       <concept_significance>500</concept_significance>
%       </concept>
%   <concept>
%       <concept_id>10002951.10003227.10003447</concept_id>
%       <concept_desc>Information systems~Computational advertising</concept_desc>
%       <concept_significance>500</concept_significance>
%       </concept>
%   <concept>
%       <concept_id>10002950.10003648.10003662.10003664</concept_id>
%       <concept_desc>Mathematics of computing~Bayesian computation</concept_desc>
%       <concept_significance>500</concept_significance>
%       </concept>
%\end{CCSXML}
%
%\ccsdesc[500]{Theory of computation~Probabilistic computation}
%\ccsdesc[500]{Information systems~Computational advertising}
%\ccsdesc[500]{Mathematics of computing~Bayesian computation}



\maketitle

\section{Introduction}
\label{intro}
Marketing, as an essential growth driver, accounts for substantial investment at many companies. Given these large investments, it is not surprising that understanding the return and optimizing the allocation of marketing investment is of foundational importance to marketing practitioners. For many decades Marketing Mix Model (a.k.a. MMM) has been leveraged as one of the most important tools in marketers’ arsenal to address such needs. Recent consumer privacy initiatives (e.g., Apple’s announcement of no-IDFA \footnote{https://developer.apple.com/app-store/user-privacy-and-data-use/} in iOS 14) further underscore the strategic importance of future-proofing marketing measurement strategies with MMM.
 
While randomized experiments \citep{vaver2011measuring} and causal models \citep{imbens2015causal} are often used for causal inference, they can be either costly  or simply infeasible \citep{lewis2015unfavorable} under some circumstances. As an alternative, MMM offers a solution by leveraging aggregated time-series data and regression to quantify the relationship between marketing and demand \citep{mccarthy1978basic}. MMM could be further tailored and enhanced for different requirements and purposes such as controlling for seasonality, trend, and other control factors \citep{larsen_2018} and introducing the geo-level hierarchy \citep{sun2017geo}. More importantly, the primary use case of MMM is often not to predict the sales, but rather to quantify the marginal effects of the different marketing tactics.

There are various issues and challenges that have to be accounted for when building MMM \citep{chan2017challenges}. First, advertising media is evolving at a fast pace so it requires modelers to take into account new marketing levers constantly. It ends up forcing modelers to face the ``small $n$ large $p$'' problem. Second, in order to have actionable insights, modelers tend to pick a high level of data granularity. However, higher level of data granularity may lead to sparse observations and outliers. Practitioners need to strike a balance between the limited amount of reliable historical data and a proper level of data granularity.  Third,  the sequential nature of the data makes it more susceptible to correlated errors, which violates the basic model assumption of ordinary least squares \citep{rawlings2001applied}.  Fourth,  there are severe endogeneity and multicollinearity concerns due to common marketing planning practices and media dynamics. For instance, setting marketing budget as percent of expected revenue is widely used, which contributes to both endogeneity as well as multicollinearity (i.e., highly correlated channel-level spend) in the models. Self-selection bias especially for the demand-capturing channels such as branded paid search \citep{blake2015consumer} can also lead to inflated measurement results if not properly addressed.  Fifth, in practice MMM usually involves a large amount of investment and a diverse set of stakeholders with whom alignments need to be secured. As such the bar for model interpretability is very high. Lastly,  it is often hard to rely on traditional machine learning approaches such as cross-validation when tuning parameters and choosing models for MMM. Modelers cannot just fall back on that, since there is rarely enough data and/or the holdout periods may not be representative of the challenging series to forecast.

It has been a long journey to build an in-house MMM solution from zero to one at Uber, which takes collaborative efforts across marketers, engineers and data scientists. Throughout this journey, we seek to address all of the above challenges. The preferred modeling solution needs to have the capability of deriving time-varying elasticity along with other temporal patterns on observational studies. More importantly, randomized experimentation results, which are generally deemed as the golden standard in measuring causality,  should be incorporated to calibrate the marginal effects in marketing levers. Benefits from both experimentation and regression modeling can be maximized when combined into one holistic framework. 

In this paper, we introduce a class of Bayesian time varying coefficient (BTVC) models that power Uber’s MMM solution. Our work brings the ideas of Bayesian modeling and kernel regression together. The Bayesian framework allows a natural way to incorporate experimentation results, and understand the uncertainty of measurement for different marketing levers. The kernel regression is used to produce the time-varying coefficients to capture the dynamics of marketing levers in an efficient and robust way.

The remainder of this paper is organized as follows. In \autoref{problem}, we describe the problem formulation and the related work. In \autoref{method}, we discuss the proposed modeling framework with emphasis on applications to MMM. In \autoref{res}, simulations as well as real-case benchmark studies are presented.  In \autoref{arch}, we talk about how to deploy the proposed models using Uber’s modern machine learning platform. \autoref{concl} is about the conclusion.

\section{Problem Formulation}
\label{problem}
\subsection{Basic Marketing Mix Model}
Expressing sales as a function of spending variables with diminishing marginal returns \citep{farris2015marketing} is one of the fundamental properties in an attribution or marketing response model. In practice, \citet{han_2020}, and \citet{lewis2018incrementality} adopted a similar strategy for their budget allocation, bidding, and attribution. In view of that, our model can be expressed in a multiplicative format as below

\begin{equation}
\label{eq-mmm}
\hat{y}_t = g(t) \cdot \prod_{p=1}^{P} f_{t,p} (x_{t,p}), ~ t=1,\cdots,T,
\end{equation}
where  $x_{t,p}$ are the regressors (i.e., the ads spending variables in our case), $\hat{y_t}$ is the marketing response, $g$ is a time-series process, $f$ is the cost curve function, $P$ is the number of regressors, and $T$ is the number of time points.  Choice of $f$ is desired to have the following properties such that
\begin{itemize}
\item $\hat{y}_t$ has an explainable structure to decompose into different driving factors,
\item temporal effects such as trend and seasonality of $\hat{y}_t$ are captured,
\item $f_{t,p}$ is differentiable and monotonically increasing,
\item $\hat{y}_t$ has diminishing marginal returns with respect to $x_{t,p}$.

\end{itemize}

\autoref{eq-mmm}  has an intuitive form as
\begin{equation}
\label{eq-mmm-2}
\hat{y}_t = e^{l_t} \cdot e^{s_t} \cdot \prod_{p=1}^{P} {x_{t,p}} ^ {\beta_{t,p}}, ~ 0 \leq \beta_{t,p} \leq 1 , ~ \forall t, p,    
\end{equation}
where $e^{l_t}$ is the trend component, $e^{s_t}$ is the seasonality, and $\beta_{t,p}$ are channel-specific time-varying coefficients.

\subsection{Related Work}
With a log-log transformation, equation (2) can be re-written as

\begin{equation}
\begin{aligned}
\label{eq-mmm-3}
ln(\hat{y}_t) &= l_t + s_t + \sum_{p=1}^{P} ln(x_{t,p}) \beta_{t,p} \\
                    &=l_t+s_t+r_t, ~ t = 1,\cdots, T,
\end{aligned}
\end{equation}

A natural idea is to use state-space models such as Dynamic Linear Model (DLM) \citep{west2006bayesian} or Kalman filter \citep{durbin2012time} to solve equation (3). However, there are some caveats associated with these approaches, especially given the goal we want to achieve with MMM:

\begin{itemize}

\item DLM with Markov Chain Monte Carlo(MCMC) sampling is not efficient and could be costly especially for high-dimensional problems,  which requires sampling for a  large number of regressors and time steps.
\item Although the Kalman filter provides analytical solutions, it has limited room for further customization such as applying restrictions on coefficient signs (e.g., positive coefficient sign for marketing spend) or a t-distributed noise process that is more outlier robust.
\end{itemize}

Meanwhile, there have been parametric and non-parametric statistical methods proposed \citep{fan2008statistical}. \citet{wu2000kernel} considered a nonparametric varying coefficient regression model with longitudinal dependent variable and cross-sectional covariates. Two kernel estimators based on componentwise local least squares criteria were proposed to estimate the time varying coefficients.  \citet{li2002semiparametric} proposed a semiparametric smooth coefficient model as a useful yet flexible specification for studying a general regression relationship with time varying coefficients. It used a local least squares method with a kernel weight function to estimate the smooth coefficient function. Nonetheless, the frequentist approaches can be expensive when doing local estimates with respect to time dimension. There are also no straightforward ways to incorporate information from experimentation results. As such, we are motivated to develop a new approach to derive time varying coefficients under a Bayesian framework for our MMM applications.

\section{Methods}
\label{method}
\subsection{Time Varying Coefficient Regression}

In view of the increased complexity of regression problem in practical MMM, we propose a Bayesian Time Varying Coefficient (BTVC) model, as inspired by the Generalized Additive Models (GAM) \citep{hastie1990generalized} and kernel regression smoothing. The key idea behind BTVC is to express regression coefficients as a weighted sum of local latent variables.

First, we define a latent variable $b_{j,p}$ for the $p$-th regressor at time $t_j$, $p=1,\cdots,P$, $j=1,\cdots,J$, $t_j \in \{1,\cdots,T\}$. There are $J$ latent variables in total for each regressor. From the perspective of spline regression, $b_{j,p}$ can be viewed as a  knot distributed at time $t_j$ for a regressor. $w$ is a time-based weighting function such that
\begin{equation}
\label{eq-beta}
\beta_{t,p} = \sum_{j} w_j(t) \cdot b_{j, p},
\end{equation}					

It is intuitive to use a weighting function taking into account the time distance between $t_j$ and $t$, 
\begin{equation}
\label{eq-kernel}
w_j(t) = k(t, t_j)/\sum_{i=1}^{J} k(t, t_i),
\end{equation}		
where $k(\cdot, \cdot)$  is the kernel function, and the denominator is to normalize the weights across knots. In practice, we have different choices for the kernel functions, such as Gaussian kernel, quadratic kernel or any other custom kernels. In \autoref{kernel}, we will discuss this in more detail.

We can also rewrite \autoref{eq-beta}  into a matrix form
\begin{equation}
\label{eq-beta-matrix}
\beta=Kb,                                                   
\end{equation}
where $\beta$ is the $T\times P$ coefficient matrix with entries $\beta_{t,j}$, $K$ is the $T\times J$ kernel matrix with normalized weight  entries $w_j(t)$, and $b$ is the $J\times P$ knot matrix with entries %$b_{j,p}$.   At time point $t$, the regression component

\begin{equation}
\begin{aligned}
\label{eq-reg}
r_t &= X_{t}\beta_{t}^T,
\end{aligned}
\end{equation}
where $\beta_t=(\beta_{t,1},\cdots,\beta_{t,p})$ and $X_t$ is the $t$-th row of regressor covariate matrix.

Besides the regression component, we can also apply \autoref{eq-beta} to other components such as trend and seasonality in \autoref{eq-mmm-3}. 
Specifically,   for the trend component,
\begin{equation}
\begin{aligned}
\label{eq-lev}
\beta_{\text{lev}} &= K_{\text{lev}}b_{\text{lev}},\\
l_t  &= \beta_{t, \text{lev}}.
\end{aligned}
\end{equation}
The trend component can be viewed as a dynamic intercept.
For the seasonality component,
\begin{equation}
\begin{aligned}
\label{eq-seas}
\beta_{\text{seas}} &= K_{\text{seas}}b_{\text{seas}},\\
s_t  &= X_{t,  \text{seas}}\beta_{t, \text{seas}}^T.
\end{aligned}
\end{equation}
$X_{t, \text{seas}}$ is the $t$-th row of seasonality covariate matrix derived from Fourier series.  In \autoref{seas}, we will discuss the seasonality in more detail.

Instead of estimating the local knots directly (i.e., $b$,  $b_\text{lev}$,  and $b_\text{seas}$ in the above equations) by optimizing an objective function, we introduce the Bayesian framework along with customizable priors to conduct the posterior sampling.

\subsection{Bayesian Framework}
\label{bayes-frame}
To capture the sequential dynamics and cyclical patterns, we use the Laplace prior to model adjacent knots

\begin{equation}
\begin{aligned}
% b_{0,\text{lev}} &\sim \text{Laplace}(0,  \sigma^{0}_\text{lev} )\\
%  b_{0,\text{seas}} &\sim \text{Laplace}(0,  \sigma^{0}_\text{seas} )\\
 b_{j,\text{lev}} &\sim \text{Laplace}(b_{j-1,\text{lev}},  \sigma_\text{lev} ),\\
b_{j,\text{seas}}& \sim \text{Laplace}(b_{j-1,\text{seas}},  \sigma_\text{seas}).
\end{aligned}
\end{equation}
The initial values ($b_{0,\text{lev}}$ and $b_{0,\text{seas}}$) can be sampled from a Laplace distribution with mean 0.  A similar approach can be found in models implemented in Facebook's Prophet package \citep{taylor2018forecasting}, which uses Laplace prior to model adjacent change points of the trend component. 

For the regression component, we introduce a two-layer hierarchy for more robust sampling due to the sparsity in the channel spending data,
\begin{equation}
\begin{aligned}
\mu_\text{reg}& \sim \mathcal{N}^+(\mu_\text{pool}, \sigma^2_\text{pool}),\\
b_\text{reg} &\sim \mathcal{N}^+(\mu_\text{reg},  \sigma^2_\text{reg}),
\end{aligned}
\end{equation}
where the superscript $+$ means a folded normal distribution (positive restriction on the coefficient signs).

In the hierarchy, the latent variable $\mu_\text{reg}$ depicts the overall mean of a set of knots of a single marketing lever. We can treat this as the overall estimate of a channel coefficient across time. This provides two favorable behaviors for the model: 
during a period with absence of spending for a channel, coefficient knot estimation of such a channel exhibits a shrinkage effect towards the overall estimate; it helps fight against over-fitting due to the volatile local structure.

\begin{figure}[ht]
\vskip 0.2in
\begin{center}
\centerline{\includegraphics[width=\columnwidth]{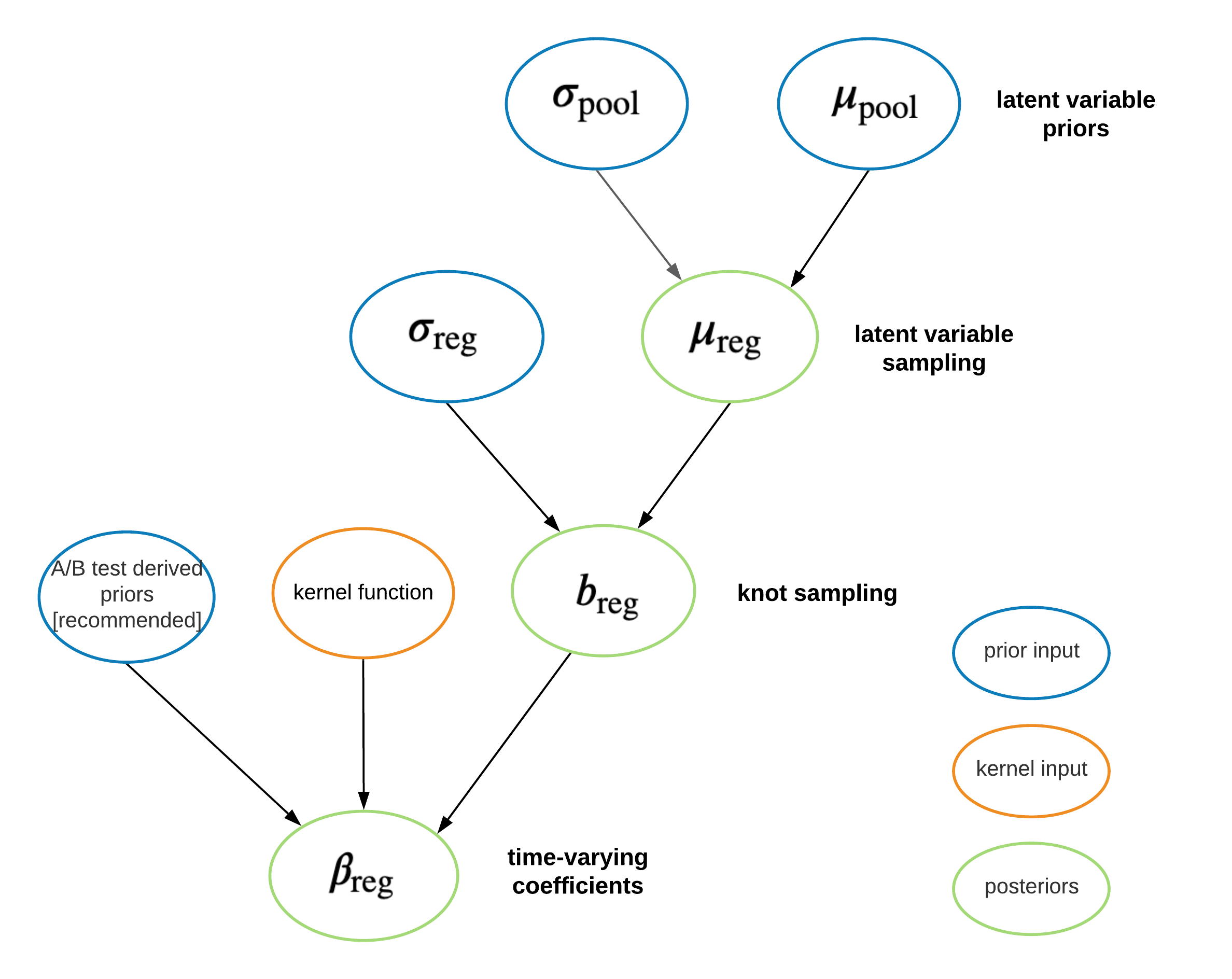}}
\caption{BTVC model flowchart.  Blue boxes present the prior related input,  where priors derived from lift tests for specific channels can be readily ingested in the framework.  Orange box represents the kernel function in use.  Green boxes represent the sampled posteriors,  which are also the quantities of interest. }
\label{ktr-flow}
\end{center}
\vskip -0.2in
\end{figure}

The two-layer hierarchy is widely adopted in hierarchical Bayesian models and the shrinkage property is sometimes called the pooling effect on regression coefficients \citep{gelman2013bayesian}. \autoref{ktr-flow} depicts the model flowchart of BTVC.  Stochastic Variational Inference \citep{hoffman2013stochastic} is used to estimate the knot coefficient posteriors from which time varying coefficient estimates can be derived using the formulas above in \autoref{eq-reg}, \autoref{eq-lev}, and \autoref{eq-seas}.

\subsection{Kernel Selection}
\label{kernel}
For the kernel function used for trend and seasonality, we propose a customized kernel, i.e.,

\noindent when $t_{i} \le t  \le t_{i+1}$ and $j\in \{i, i+1\}$,
\begin{equation}
\begin{aligned}
k_\text{lev}(t,  t_j) &= 1 - \frac{|t - t_j|}{t_{i+1} - t_i};\\
\end{aligned}
\end{equation}
otherwise zero values are assigned. This kernel bears some similarity with the triangular kernel.

For the kernel function used for regression, we adopt the Gaussian kernel, i.e.,
\begin{align}
k_\text{reg}(t, t_j;\rho) = \exp \left( -\frac{(t-t_j)^2}{2\rho^2}  \right),
\end{align}
where $\rho$ is the scale parameter. Other kernels such as Epanechnikov kernel and quadratic kernel etc., could also be leveraged for the regression component.

\subsection{Seasonality}
\label{seas}
Seasonality is a pattern that repeats over a regular period in a time series. To estimate seasonality, a standard approach is to decompose time-series into trend, seasonality and irregular components using  Fourier analysis \citep{de2011forecasting}. This method represents the time series by a set of elementary functions called basis such that all functions under study can be written as linear combinations of the elementary functions in the basis. These elementary functions involve the sine and cosine functions or complex exponential. The Fourier series approach describes the fluctuation of time series in terms of sinusoidal behavior at various frequencies.

Specifically, for a given period $S$ and a given order $k$, two series $cos(2k\pi t/S)$ and $sin(2k\pi t/S)$ will be generated to capture the seasonality. For example, with daily data, $S=7$ represents the weekly seasonality, while $S=365.25$ represents the yearly seasonality.

%\subsection{Improvements from Traditional MMM}
%With the modeling approach presented in this section, many limitations and challenges could be addressed as mentioned in the introduction section, including model uncertainty, selection bias, and expert knowledge ingestion, etc.
%
%\begin{table}[t]
%%\vskip 0.15in
%\begin{center}
%\begin{tiny}
%\begin{tabular}{ll} \hline
%Model & How they are addressed in BTVC  \\ \hline
%
%Dynamic measurement of marketing levers & Time-varying regression coefficients \\
%Uncertainty of marketing levers & Bayesian inference \\
%Ingestion of lift test results, especially for the endogenous channel to avoid selection bias &
%Incorporate experimentation insights as priors in the Bayesian model \\
%Outlier robustness &t-distributed noise process\\
%Data sparsity&Shrink to the overall estimate\\
%Multicollinearity&Coefficient size shrinkage by Bayesian priors\\\hline
%
%\end{tabular}
%\caption{MMM challenges and how they are addressed by BTVC.}
%\label{address}
%\end{tiny}
%\end{center}
%\vskip -0.1in
%\end{table}

\section{Results}
\label{res}
\subsection{Simulations}
\subsubsection{Coefficient Curve Fitting}
\label{sec-curve-fit}

We conduct a simulation study based on the following model
$$y_t = \text{trend} + \beta_{1t}x_{1t} + \beta_{2t}x_{2t} + \beta_{3t}x_{3t} + \epsilon_t,~ t = 1,\cdots T,$$
where the trend and $\beta_{1t}, \beta_{2t}, \beta_{3t}$ are all random walks. The covariates         
$x_{1t}, x_{2t}, x_{3t}\sim \mathcal{N}(3, 1)$ are independent of the error term $\epsilon_t \sim \mathcal{N}(0, .3)$.

\begin{figure}[ht]
\vskip 0.2in
\begin{center}
\centerline{\includegraphics[width=\columnwidth]{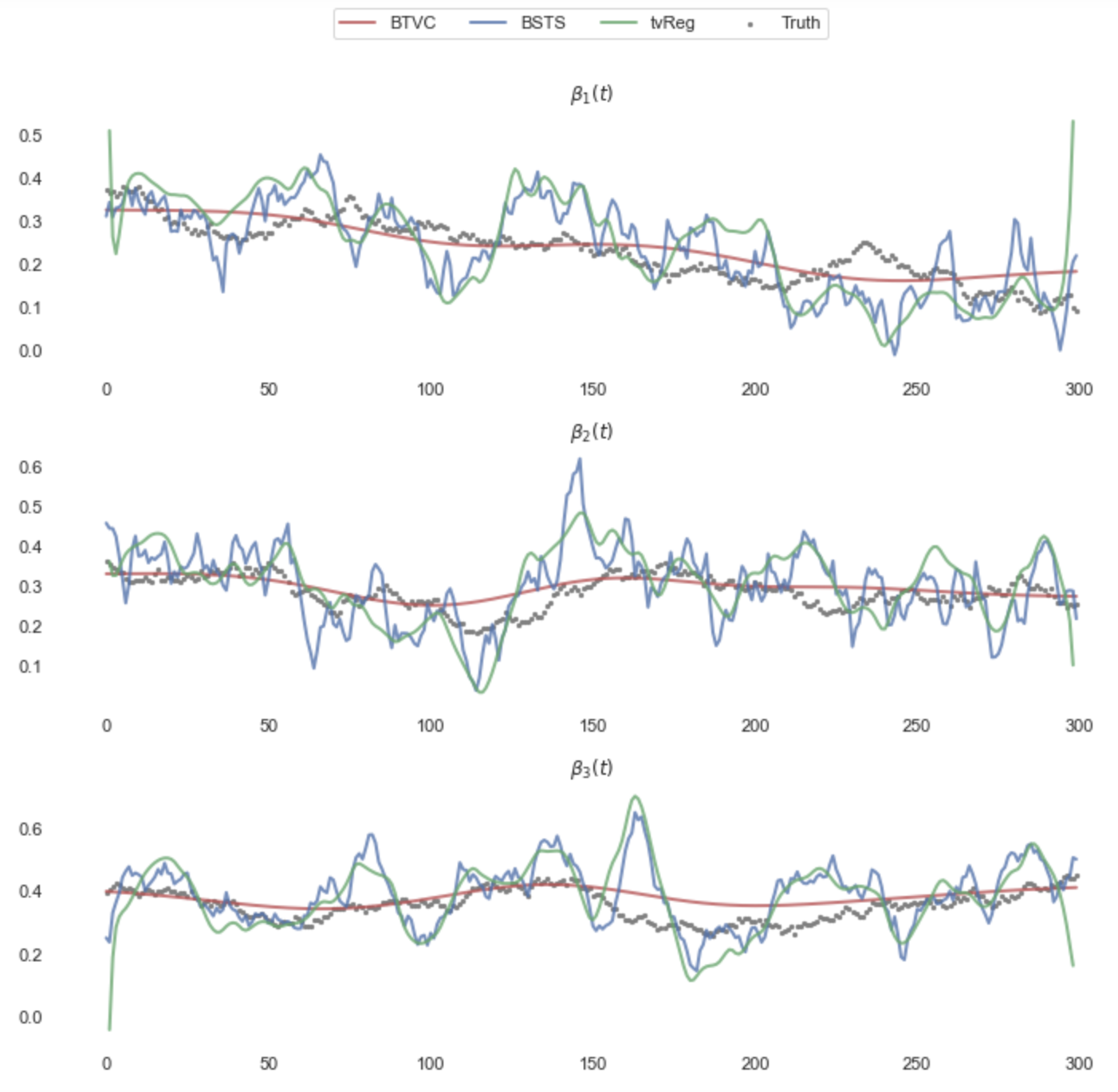}}
\caption{Comparison of the BSTS, tvReg and BTVC estimates of coefficient functions. The true values are plotted in grey dots, the blue line is BSTS estimate, the green line refers to tvReg and the red line for BTVC.}
\label{curve-fit}
\end{center}
\vskip -0.2in
\end{figure}

In our study, we compare BTVC with two other time varying regression models available in R CRAN: Bayesian structural time series (BSTS) \citep{scott2014predicting}, and time varying coefficient for single and multi-equation regressions (tvReg) \citep{casas2021tvreg}.  We set $T$=300 and calculate the average Mean Squared Errors (MSE) against the truth of each regressor across 100 simulations. The estimated coefficient curves of a sample is plotted in \autoref{curve-fit}.  The results are reported in \autoref{tbl-curve-fit}, which demonstrates  that  BTVC has a better accuracy on the coefficient estimation over the other two models in consideration.

\begin{table}[t]
\begin{center}
\begin{tabular}{lcccr} \hline
Model & $\beta_1(t)$  & $\beta_2(t)$ & $\beta_3(t)$  \\ \hline
BSTS &0.0067& 0.0078 & 0.0080  \\
tvReg &  0.0103& 0.0103 & 0.0096\\
BTVC & 0.0030& 0.0026  & 0.0029  \\ \hline
\end{tabular}
\caption{Average of mean squared errors based on 100 times simulations.}
\label{tbl-curve-fit}
\end{center}
\end{table}

\subsubsection{Experimentation Calibration}
\label{exp-calib}
One appealing property of the BTVC model is its flexibility to ingest any experimentation based priors for any regressors (e.g., advertising channels) since experiments are often deemed as a trustworthy source to tackle the challenges as mentioned in \autoref{intro}.

To illustrate this feature of BTVC, we first fit a BTVC model on the simulated data, which are generated using a similar simulation scheme as outlined in \autoref{sec-curve-fit}.  Next, we assume there is one lift test for the first and third regressors, respectively, and two lift tests for the second regressor. All the tests have a 30-step duration. We use the simulated values as the ``truth'' derived from the tests, and ingest them as priors into BTVC models. The results are summarized in Figure 3.  As expected, the credible intervals during the ingestion periods and the adjacent neighborhood become narrower, compared to the ones without prior knowledge. Moreover, with this calibration, the coefficient curves are more aligned with the truth around the neighborhood of the test ingestion period. To demonstrate this, in \autoref{tbl-beta-fit} we calculated the symmetric mean absolute percentage error (SMAPE) and pinball loss (with 2.5\% and 97.5\% target quantiles) between the truth and the estimation for the following 30 steps after the prior ingestion period. 

\begin{figure}[ht]
\begin{center}
\begin{subfigure}{.5\textwidth}
  \centering
  \includegraphics[width=\columnwidth]{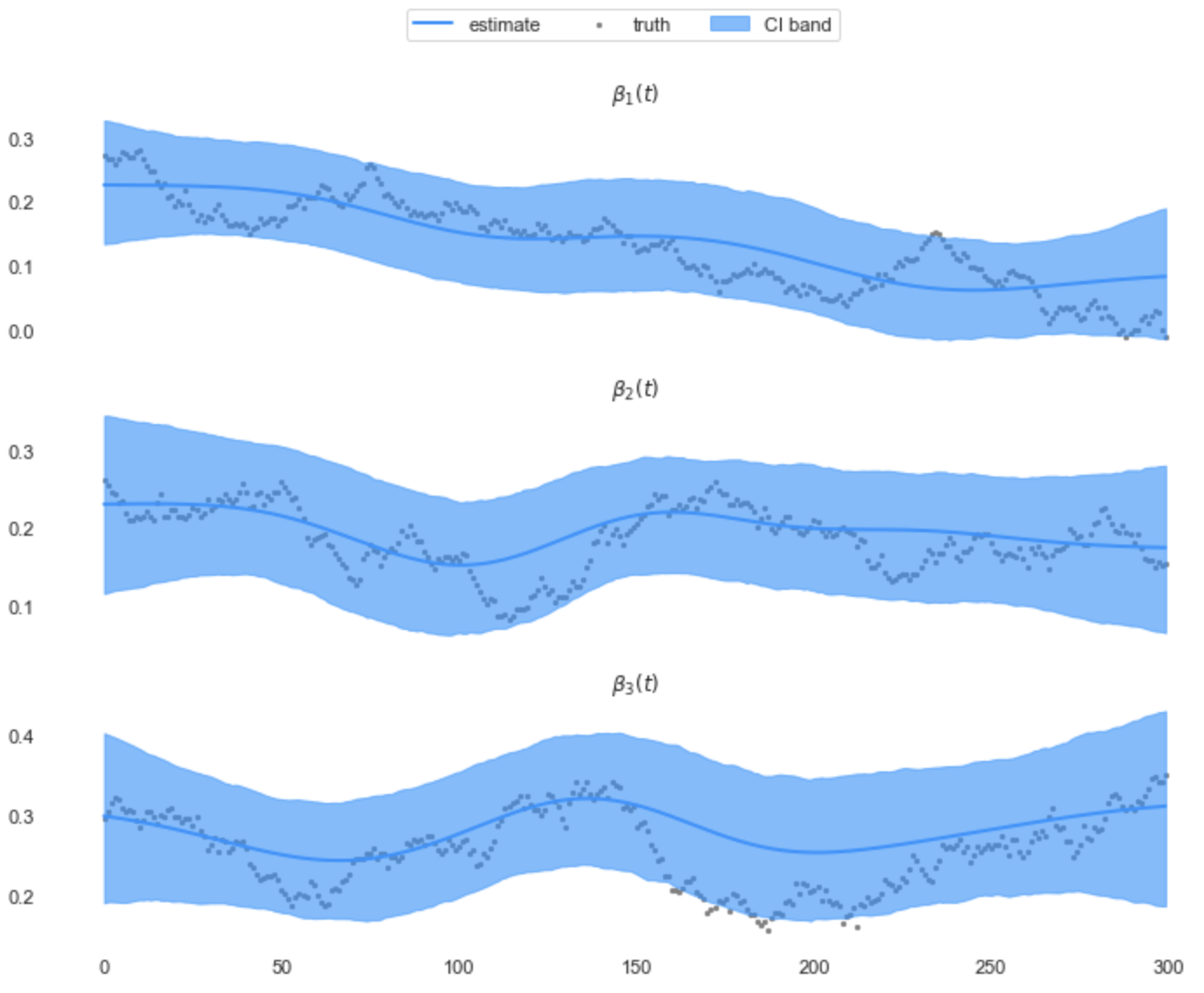}
  \caption{without prior ingestion}
\end{subfigure}

\begin{subfigure}{.5\textwidth}
  \centering
  \includegraphics[width=\columnwidth]{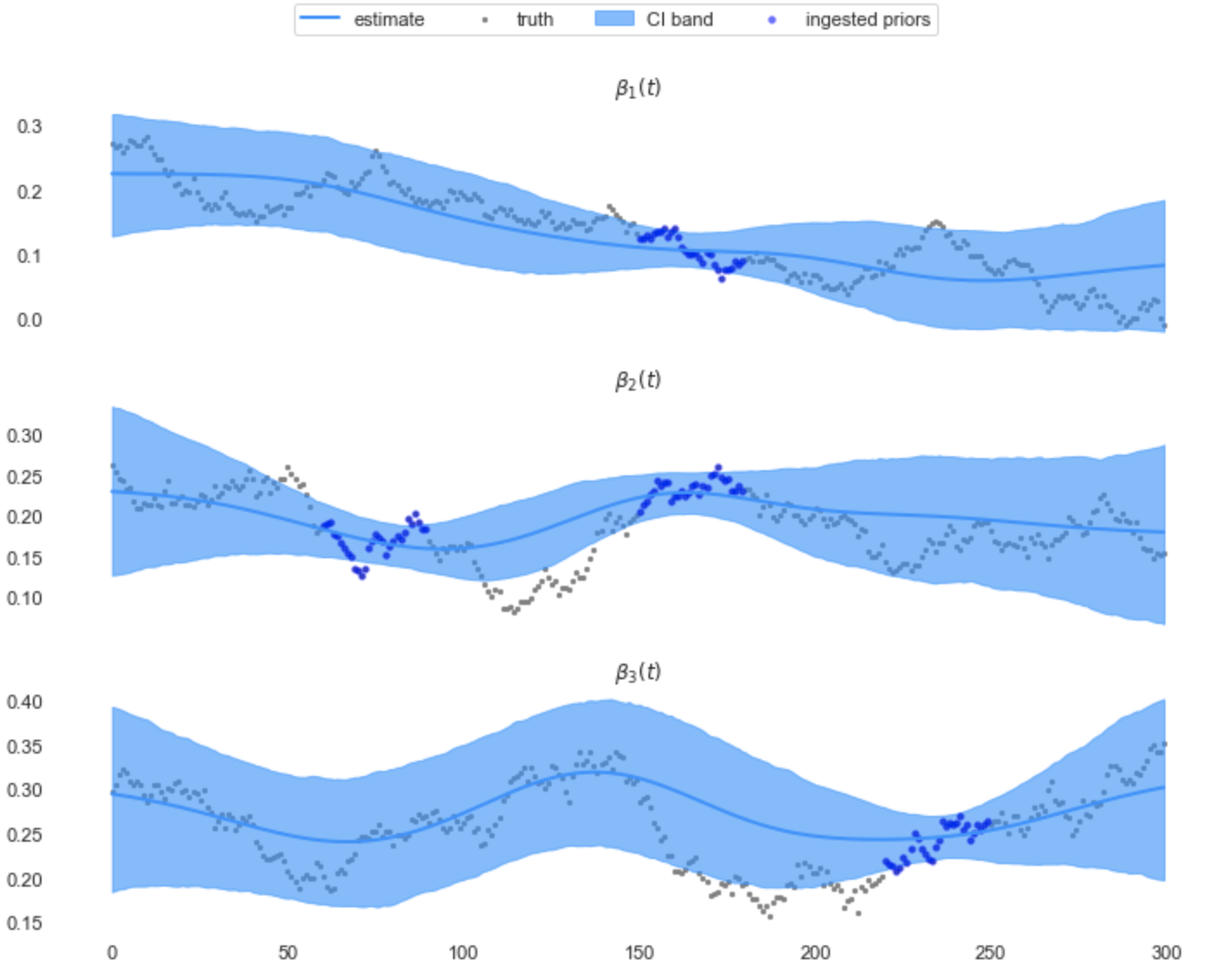}
  \caption{with prior ingestion}
\end{subfigure}

\caption{ (a) The coefficient estimation of BTVC mode without the prior ingestion.  (b) The coefficient estimation of BTVC model with the prior ingestion. The test ingestion periods are highlighted in blue dots. The black solid lines are the simulated truth, while the red ones are the estimations. The shaded bands represent the 95\% credible intervals}
\end{center}
\label{beta-fit}
\end{figure}

\begin{table}[t]
\begin{center}
\begin{tabular}{lc|c|cc|cr} \hline
&\multicolumn{2}{c|}{SMAPE} & \multicolumn{4}{c}{Pinball Loss}\\ \hline
 & \multirow{2}{*}{w/o priors}  & \multirow{2}{*}{w priors} &  \multicolumn{2}{c|}{w/o priors} & \multicolumn{2}{c}{w priors} \\ 
 &   &  & lower & upper & lower & upper\\ \hline
$\beta_1(t)$ &0.39& 0.21 & 0.0009 & 0.0032 & 0.0005 & 0.0019  \\
$\beta_2(t)$ &  1.37& 1.25 & 0.0021&0. 0019& 0.0011&0.0014\\
$\beta_3(t)$ &0.30&0.18  &   0.0017& 0.0028& 0.0014&0.0013\\ \hline
\end{tabular}
\caption{SMAPE and pinball loss of coefficient estimates for models without and with prior ingestions. The metrics are calculated using the 30-step coefficients following the test ingestion period.  Lower (2.5\%) and upper(97.5\%) quantiles are reported for pinball loss. With prior ingestion, the coefficient estimation accuracy is improved significantly.}
\label{tbl-beta-fit}
\end{center}
\end{table}

\subsubsection{Shrinkage Property}

\begin{figure}[ht]
\vskip 0.2in
\begin{center}
\centerline{\includegraphics[width=\columnwidth]{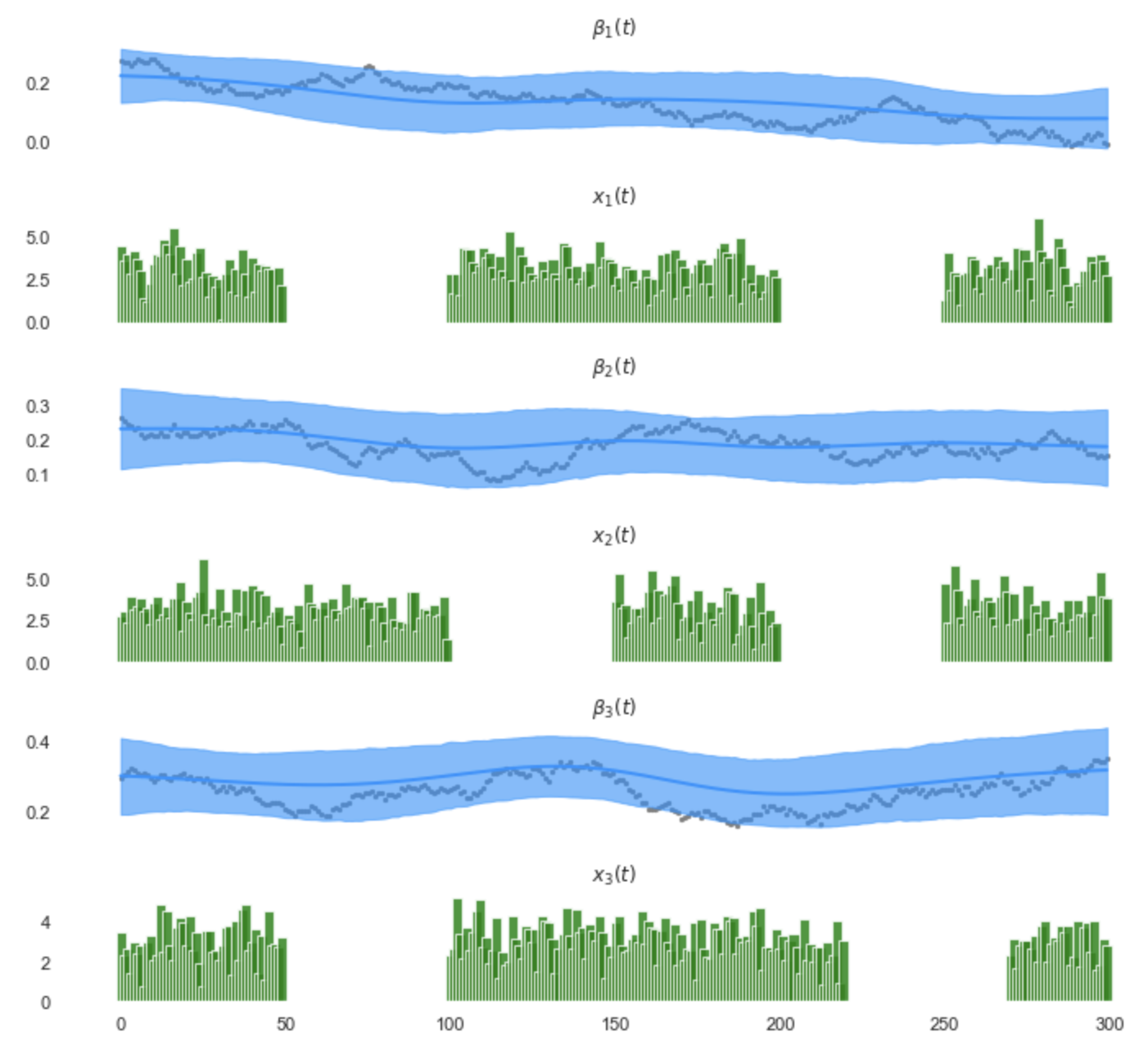}}
\caption{Coefficient estimation exhibits a shrinkage effect towards the grand mean of the coefficient curve instead of zero. The histograms in green represent the covariates.}
\label{sparse}
\end{center}
\vskip -0.2in
\end{figure}

In real-life MMM data, it is common to observe an intermittent marketing spending  pattern, which means that a given advertising channel is characterized by many minuscule or zero spends over time.  In BTVC, the coefficient estimation over such sparse periods exhibits a shrinkage effect towards the grand mean of the coefficient curve instead of zero, which is due to the hierarchical structure of the model as discussed in \autoref{bayes-frame}.    \autoref{sparse} is a simulation example to demonstrate this property, where the covariates are plotted along with the estimated coefficients. 

\subsection{Real Case Studies}
\subsubsection{Forecasting Benchmark}
To benchmark the model's forecasting accuracy,  we conduct a real-case study using Uber Eats’ data across 10 major countries or markets. Each country series consists of the daily number of first orders in Uber Eats by newly acquired users. The data range spans from Jan 2018 to Jan 2021 including a typical Covid-19 period. The scheme change caused by Covid-19 poses a big challenge for modeling.

We compared BTVC with two other time series modeling techniques, SARIMA \citep{seabold2010statsmodels} and Facebook Prophet \citep{taylor2018forecasting}. Both Prophet and BTVC models use Maximum A Posterior (MAP) estimates and they are configured as similarly as possible in terms of optimization and seasonality settings. For SARIMA, we fit the $(1, 1, 1) \times (1, 0, 0)_S$ structure by maximum likelihood estimation (MLE) where $S$ represents the choice of 
seasonality. In our case, $S = 7$ for the weekly seasonality.

We use SMAPE as the performance benchmark metric
$$\text{SMAPE} = \sum^{h}_{t=1} \frac{|F_t - A_t|}{(|F_t| + |A_t|)/2}, $$
where $F_t$ (predicted) or $A_t$ (actual) represents the value measured at time $t$ and $h$ is the forecast horizon which can also be considered as the ``holdout'' length in a backtesting process.  We used $h=28$ as the forecast horizon with 6 splits (i.e., 6 different cuts of the data with incremental training length) in this exercise.

\begin{figure}[ht]
\vskip 0.2in
\begin{center}
\centerline{\includegraphics[width=\columnwidth]{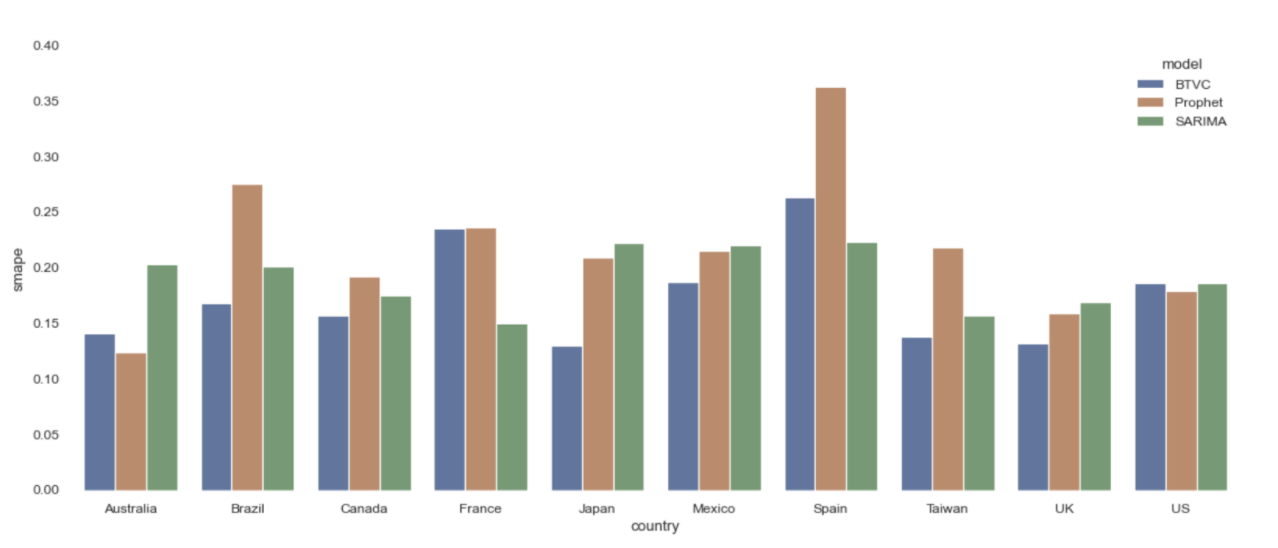}}
\caption{Bar plots of SMAPE on 10 countries with Uber Eats data.Blue bars represent the results from BTVC, orange ones Prophet, and yellow ones SARIMA.}
\label{bar-eater}
\end{center}
\vskip -0.2in
\end{figure}

\autoref{bar-eater} depicts the SMAPE results across the 10 countries, and \autoref{tbl-eater} gives the average and standard deviation of SMAPE values for the three models in consideration. As a result, BTVC outperforms the other two models for the majority of 10 countries in terms of SMAPE.

\begin{table}[t]
%\vskip 0.15in
\begin{center}
\begin{tabular}{lcccr} \hline
Model & Mean of SMAPE & Std of SMAPE  \\ \hline
SARIMA & 0.191 & 0.027 \\
Prophet & 0.218 & 0.067\\
BTVC & 0.174 & 0.045 \\ \hline
\end{tabular}
\caption{SMAPE comparison across models. It shows that BTVC outperforms the other two models in terms of average SMAPE across 10 countries.}
\label{tbl-eater}
\end{center}
\end{table}

\subsubsection{Attribution}
Simulations in \autoref{exp-calib} show that ingestion of multiple lift test results can help  improve estimation of marketing incrementality. In this section, we use real marketing data to further demonstrate the benefits of MMM calibration with experimentation. Simply put, properly designed and executed experimentation provides the ground truth for marketing measurement. The  more valid experiments one can ingest for the measurement of a given marketing lever, the more accuracy one can achieve in terms of understanding its marginal impact, despite experimentation-based insights not being available all the time.  BTVC provides a rigorous algorithm that enables informing extrapolation of channel-level elasticity for periods with no experimentation coverage using all available tests, while simultaneously adapting to the new observational data as well as taking into account the differential statistical strength and temporal differences of various tests. 

In this real case study, we leverage data and experimentation insights for a given paid channel where three usable experiments are available, covering different but temporally adjacent time periods, i.e., Experiment-1, Experiment-2, Experiment-3. In terms of temporal order, Experiment-1 is the oldest experiment while Experiment-3 is the most recent one.

We construct two attribution validation studies, each addressing a real-life scenario:
\begin{itemize}
\item Study 1: Treat Experiment-2 as the unobserved truth while using lift insights from Experiment-1 and Experiment-3 for MMM calibration.  This test addresses the scenario where a given paid channel has some experiments but there are non-trivial gaps in between. 
\item Study 2: Treat Experiment-3  as the unobserved truth while using lift insights from Experiment-1 and Experiment-2 for MMM calibration. This test addresses the scenario where a given paid channel has some old experiments but there are no new experiments planned or the new experiments will take a long time to complete.
\end{itemize}

For both studies,  the following models are fitted:
\begin{itemize}
\item Baseline Model 1:  No experimentation-based priors will be used. 
\item Baseline Model 2:  Experimentation insights from Experiment-1 will be used.
\item Baseline Model 3: Experimentation insights from Experiment-2 will be used (in Study 2 only).
\item Champion Model:  Experimentation insights from the periods with known experimentation will be ingested as Bayesian priors. Specifically,  for Study 1 the insights from Experiment-1 and Experiment-3 will be ingested as priors, while for Study 2 insights from Experiment-1 and Experiment-2 will be ingested as priors.
\end{itemize}

\begin{table}[t]
%\vskip 0.15in
\begin{center}
\begin{tabular}{lccccc} \hline
\multirow{2}{*}{Scenario}  & Experiment & Baseline & Baseline &Baseline& Champion \\ 
& Attribution &Model 1 & Model 2&Model 3&Model \\\hline
Study 1& {1889} &{1427} &{1556} &N/A&{1717} \\
Study 2& {1147} &{784} &{855} & {1311} & {1240} \\\hline
\end{tabular}
\caption{Attributions on a paid channel from the fitted models and the ones derived from the experimentation, which can be deemed as the source of truth.}
\label{tbl-attr}
\end{center}
\end{table}

In \autoref{tbl-attr}, we provide for each study the attribution insights from various models (i.e.,  baseline as well as champion models) and the experiments serving as the unobserved truth.  By leveraging the results,  to some extent, the following hypotheses could be validated:

\begin{itemize}
\item For Study 1, attribution accuracy for the gap period in between 2 periods covered with experimentation increases if insights from both adjacent tests are ingested,  as attribution based on the champion model (i.e., 1717) is closest to experiment-based attribution (i.e., 1889). 

\item For Study 2, attribution accuracy for the most recent period with no test coverage is the best when all available older tests are used for model calibration, as attribution from the champion model (i.e., 1240) is closest to experiment-based attribution (i.e., 1147).  It also demonstrates that it is helpful to leverage older test in addition to the most recent one,  as attribution based on the champion model (i.e., 1240) is closer to experiment-based attribution (i.e., 1147) than baseline model 3 (i.e., 1311).
\end{itemize}

In summary,  ingestion of multiple experiments lead to significant improvement in attribution accuracy. Most notably, the validation results indicate that:  contrary to conventional wisdom in marketing measurement, using only the latest experiment for measurement model calibration does not lead to superior attribution accuracy for future periods. Older experimentation, albeit more temporally distant from the attribution period of interest in the future, can still add value.  BTVC enables modelers to seamlessly combine insights from new as well as old experiments, producing attribution insights with better accuracy. 

%Attributions from the champion and baseline models as well as the actual lift results from the experimentation serving as ``unobserved truth'' are summarized in \autoref{tbl-attr}. As expected,  the Champion Model outperforms both baseline models, and Baseline Model 2 outperforms Baseline Model 1 in terms of attribution accuracy for the test periods with respect to the ``unobserved truth''. 

%Moreover, we fit another model by only ingesting lift insights from Experiment-2 and then use this model to produce attribution (i.e., 1311) for the time period corresponding to Experiment-3 . Comparing attribution results from this new model to Champion Model in Study 2, we observe superior attribution accuracy for the latter, further illustrating the beneficial impact of ingesting multiple experimentation insights. 

\section{Architecture}
\label{arch}

\subsection{Implementation}

We implemented the BTVC model as a feature branch in our open sourced package {\em Orbit} \citep{ng2021orbit} by Uber. Orbit is a software package aiming to simplify time series inferences and forecasting with structural Bayesian time series models for real-world cases and research. It provides a familiar and intuitive initialize-fit-predict interface for time series tasks, while utilizing the probabilistic programming languages such as Stan \citep{carpenter2017stan} and Pyro \citep{bingham2019pyro} under the hood. 
\subsection{Deployment}
The BTVC deployment system is customized by leveraging Michelangelo \citep{hermann2017}, a machine learning (ML) platform developed by Uber.  Michelangelo provides centralized workflow management for end-to-end modeling process.With the help of Michelangelo,  BTVC deployment system is able to automate data preprocessing, model training,  validations, predictions, and measurement monitoring at scale.

\begin{figure}[ht]
\vskip 0.2in
\begin{center}
\centerline{\includegraphics[width=\columnwidth]{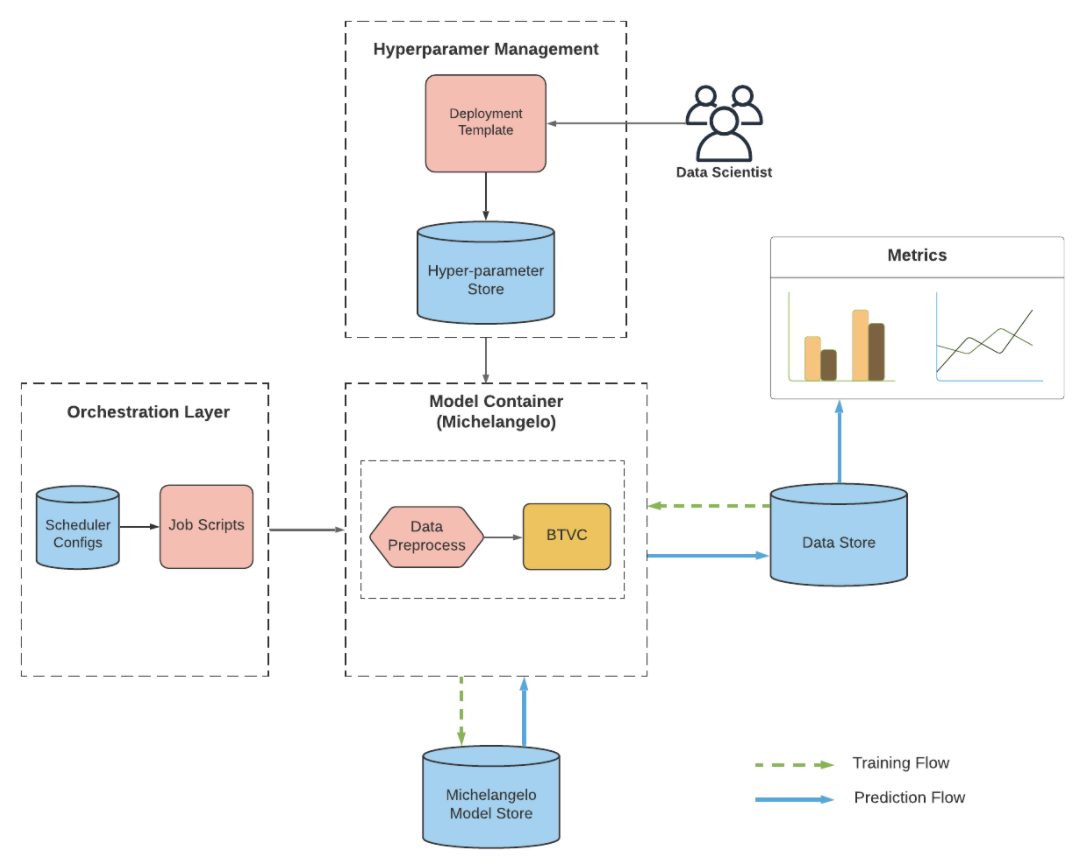}}
\caption{BTVC deployment system.}
\label{dep-flow}
\end{center}
\vskip -0.2in
\end{figure}

The deployment workflow summarized in \autoref{dep-flow} consists of three main components:
\begin{itemize}
\item Hyperparameter management layer: this is to store and manage the various supplement data needed for BTVC model training such as normalization scalar,  adstock \citep{jin2017bayesian},  lift test based priors, as well as model specific hyperparameters. 
\item Orchestration layer: this is to upload and trigger the model training job. 
\item Model container: a docker container including all the essential modeling code to be integrated with Michelangelo's ecosystem.
\end{itemize}

\section{Conclusion}
\label{concl}

In this paper, we propose a Bayesian Time Varying Coefficient (BTVC) model in particular developed for MMM applications at Uber. By assuming the local latent variables follow certain probabilistic distributions, a kernel-based smoothing technique is applied to produce the dynamic coefficients. This modeling framework entails a comprehensive solution for the challenges faced by traditional MMM.  More importantly, it enables marketers to leverage multiple experimentation results in an intuitive yet scientific way.  Simulations and real case benchmark studies demonstrate BTVC’s superiority in prediction accuracy and flexibility in experimentation ingestion. We also present the model deployment system, which can serve model training and predictions in real time without human oversight or intervention in a scalable way.

\section{Acknowledgments}

The authors would like to thank Sharon Shen, Qin Chen, Ruyi Ding, Vincent Pham, and Ariel Jiang for their help on this project, Dirk Beyer,  and Kim Larsen for their comments on this paper.

\bibliographystyle{ACM-Reference-Format}
\bibliography{arxiv-ktr}

\end{document}